\documentclass[twocolumn,showpacs,amsmath,amssymb,10pt]{revtex4}
\usepackage{amsfonts}
\usepackage{bm}

\newcommand{\ot}{{\,\otimes\,}}
\newcommand{{\Cd}}{{\mathbb{C}^d}}
\newcommand{\sbsigma}{{\mbox{\scriptsize \boldmath $\sigma$}}}
\newcommand{\sbalpha}{{\mbox{\scriptsize \boldmath $\alpha$}}}
\newcommand{\sbbeta}{{\mbox{\scriptsize \boldmath $\beta$}}}
\newcommand{\bsigma}{{\mbox{ \boldmath $\sigma$}}}
\newcommand{\balpha}{{\mbox{ \boldmath $\alpha$}}}
\newcommand{\bgamma}{{\mbox{ \boldmath $\gamma$}}}
\newcommand{\sbgamma}{{\mbox{\scriptsize \boldmath $\gamma$}}}

\newcommand{\bbeta}{{\mbox{ \boldmath $\beta$}}}
\newcommand{\bmu}{{\mbox{ \boldmath $\mu$}}}
\newcommand{\bnu}{{\mbox{ \boldmath $\nu$}}}
\newcommand{\sbmu}{{\mbox{\scriptsize \boldmath $\mu$}}}
\newcommand{\sbnu}{{\mbox{\scriptsize \boldmath $\nu$}}}
\def\oper{{\mathchoice{\rm 1\mskip-4mu l}{\rm 1\mskip-4mu l}%
{\rm 1\mskip-4.5mu l}{\rm 1\mskip-5mu l}}}
\def\<{\langle}
\def\>{\rangle}
\newtheorem{theorem}{Theorem}
\newtheorem{proposition}{Proposition}

\begin{document}
\title{\textbf{Multipartite circulant states with
positive partial transposes}}
\author{Dariusz
Chru\'sci\'nski\thanks{email: darch@phys.uni.torun.pl}$\ $ and
Andrzej
Kossakowski \\
Institute of Physics, Nicolaus Copernicus University,\\
Grudzi\c{a}dzka 5/7, 87--100 Toru\'n, Poland}

\begin{abstract}
We construct a large  class of multipartite qudit  states which
are positive under the family of partial transpositions. The
construction is based on certain direct sum decomposition of the
total Hilbert space displaying characteristic circular structure
and hence generalizes a class of bipartite  circulant states
proposed recently by the authors.  This  class contains many well
known examples of multipartite quantum states from the literature
and gives rise to a huge family of completely new states.

\end{abstract}

\maketitle

\section{Introduction}

Quantum entanglement is one of the most remarkable features of
quantum mechanics and it leads to powerful applications like
quantum cryptography, dense coding and quantum computing
\cite{QIT,Horodecki-review}.

It is well known that it is very hard to check whether a given
density matrix describing a quantum state of the composite system
is separable or entangled.  There are several operational criteria
which enable one to detect quantum entanglement (see e.g.
\cite{Horodecki-review} for the recent review). The most famous
Peres-Horodecki criterion \cite{Peres,PPT} is based on the partial
transposition: if a state $\rho$ is separable then its partial
transposition $(\oper \ot \tau)\rho$ is positive (such states are
called PPT state). The structure of this set is of primary
importance in quantum information theory. Unfortunately, this
structure is still unknown, that is, one may easily check whether
a given state is PPT but we do not know how to construct a general
quantum state with PPT property.

Recently \cite{CIRCULANT} we proposed a large class of bipartite
PPT states  which are based on certain cyclic decomposition of the
total Hilbert space $\mathbb{C}^d \ot \mathbb{C}^d$ --- we called
them circulant states. The crucial property of this class is that
a partial transposition of the circulant state has again a
circular structure corresponding to another cyclic decomposition
of $\mathbb{C}^d \ot \mathbb{C}^d$. Interestingly, many well known
examples of PPT states fit the class of circulant states
\cite{CIRCULANT}.

In the present paper we generalize the construction of circulant
states to multipartite systems --- $N$  qudits  living in
$\mathbb{C}^d \ot \ldots \ot \mathbb{C}^d$ ($N$ copies). This
space can be in a natural way decomposed into $d^{N-1}$ subspaces
of dimension $d$ or $d$ subspaces of dimension $d^{N-1}$.
Multipartite circulant state is defined as a convex combination of
positive operators supported on these orthogonal subspaces. It
turns out that the family of partial transpositions maps these
states into circulant operators supported on another family of
subspaces related by a circular multipartite structure. Again, we
show that many well known examples of multipartite states belong
to our class.

 Recently, there is a
considerable effort to explore multipartite systems
\cite{MULTI1}--\cite{I&II} and multipartite circulant states
introduced in this paper may shed new light on the more general
investigation of multipartite entanglement.

The paper is organized as follows: for pedagogical reason we first
illustrate our general method for $d=2$ in  Section
\ref{2-QUBITS}. We recall basic construction for 2 qubits from
\cite{CIRCULANT} and then analyze in details 3 qubit circulant
states. We illustrate our construction with well known examples of
3 qubit states from the literature. Then we present a general
construction of $N$ qubit states. Section \ref{N-QUDITS} discusses
circulant states of $N$ qudits. Final conclusions are collected in
the last section.

\section{$N$--qubit states} \label{2-QUBITS}

\subsection{2 qubits}

Consider a density matrix living in $\mathbb{C}^2 \ot
\mathbb{C}^2$ which is given by
\begin{equation}\label{2-qubits}
    \rho = \rho_0 +  \rho_1\ ,
\end{equation}
where $\rho_0$ and $\rho_1$ are supported on two orthogonal
subspaces
\begin{eqnarray}\label{}
    \Sigma_0 &=& \mbox{span}\left\{ e_0 \ot e_0\, , e_1 \ot e_1 \right\} \
    , \nonumber    \\
    \Sigma_1 &=& \mbox{span}\left\{ e_0 \ot e_1\, , e_1 \ot e_0 \right\} \ ,
\end{eqnarray}
and $\{e_0,e_1\}$ is a computational base in $\mathbb{C}^2$. It is
clear that $\{\Sigma_0,\Sigma_1\}$ defines the direct sum
decomposition of $\mathbb{C}^2 \ot \mathbb{C}^2$, that is
\[  \Sigma_0 \oplus \Sigma_1\, = \, \mathbb{C}^2 \ot \mathbb{C}^2\ . \]
We call it a circulant decomposition because its structure is
determined by the cyclic shift $S : \mathbb{C}^2 \rightarrow
\mathbb{C}^2$ defined by
\begin{equation}\label{Shift-2}
    {S} \, e_i = e_{i+1}\ , \ \ \ \ \ (\mbox{mod}\ 2)\ .
\end{equation}
One finds that
\begin{equation}\label{}
    \Sigma_1 = (\oper \ot S)\, \Sigma_0\ ,
\end{equation}
and hence
\begin{eqnarray}\label{ro-2-qubit}
\rho_0 &=& \sum_{i,j=0}^1\, a_{ij}\, e_{ij} \ot e_{ij}\ , \\
\rho_1 &=& \sum_{i,j=0}^1\, b_{ij}\, e_{ij} \ot S\,e_{ij} S^*\ ,
\nonumber \\ &=& (\mathbb{I} \ot S)\left( \sum_{i,j=0}^1\,
b_{ij}\, e_{ij} \ot e_{ij}\right) (\mathbb{I} \ot S)^*\ ,
\end{eqnarray}
where    $e_{ij} := |e_i\>\<e_j|$, and one adds mod 2. Now, since
$\rho_0$ and $\rho_1$ are supported on two orthogonal subspaces
$\Sigma_0$ and $\Sigma_1$ one has an obvious

\begin{proposition}
$\rho$ defined in (\ref{2-qubits}) is a density matrix iff
\begin{itemize}

\item $a=[a_{ij}]$ and $b=[b_{ij}]$ are $2 \times 2$
semi-positive matrices, and

\item ${\rm Tr}(a + b)=1\ $.

\end{itemize}
\end{proposition}

Now, the crucial observation is that partially transposed matrix
$\rho^{\tau} =(\oper \ot \tau)\rho$  belongs to the same class as
original $\rho$
\begin{equation}\label{2C-T}
\rho = \left( \begin{array}{cc|cc}
    a_{00} & \cdot & \cdot & a_{01} \\
    \cdot      & b_{00} & b_{01} & \cdot \\ \hline
    \cdot      & b_{10} & b_{11} & \cdot \\
    a_{10} & \cdot & \cdot & a_{11} \end{array} \right)\ ,
\end{equation}
and
\begin{equation}
        \rho^{\tau} = \left( \begin{array}{cc|cc}
    \widetilde{a}_{00} & \cdot & \cdot & \widetilde{a}_{01} \\
    \cdot      & \widetilde{b}_{00} & \widetilde{b}_{01} & \cdot \\ \hline
    \cdot      & \widetilde{b}_{10} & \widetilde{b}_{11} & \cdot \\
    \widetilde{a}_{10} & \cdot & \cdot & \widetilde{a}_{11} \end{array} \right)\
    ,
\end{equation}
where the matrices $\widetilde{a} = [\widetilde{a}_{ij}]$ and
$\widetilde{b} = [\widetilde{b}_{ij}]$ read as follows
\begin{equation}\label{}
    \widetilde{a} = \left( \begin{array}{cc}
    a_{00} & b_{01} \\
    b_{10} & a_{11} \end{array} \right) \ , \ \ \ \
\widetilde{b} = \left( \begin{array}{cc}
    b_{00} & a_{01} \\
    a_{10} & b_{11} \end{array} \right)\ ,
\end{equation}
that is, both $\rho$ and $\rho^{\tau}$ are circulant bipartite
operators. Therefore, one arrives at
\begin{proposition}
A circulant state represented by (\ref{2-qubits}) is {\rm PPT} iff
$\widetilde{a}=[\widetilde{a}_{ij}]$ and
$\widetilde{b}=[\widetilde{b}_{ij}]$ are $2 \times 2$
semi-positive matrices.
\end{proposition}
Note, that matrices $\widetilde{a}$ and $\widetilde{b}$ may be
rewritten in the following transparent way
\begin{equation}\label{}
    \widetilde{a} = a \circ \mathbb{I} + b \circ S\ ,
\end{equation}
and similarly
\begin{equation}\label{}
    \widetilde{b} = b \circ \mathbb{I} + a \circ S\ ,
\end{equation}
where $x \circ y$ denotes the Hadamard product of two matrices $x$
and $y$.

{\it Examples.} 1. Bell states: $|\psi^\pm\> = (|00\> \pm
|11\>)/\sqrt{2}$
\begin{equation}\label{}
|\psi^\pm\>\<\psi^\pm| = \frac 12 \left( \begin{array}{cc|cc}
    1 & \cdot & \cdot & \pm 1 \\
    \cdot      & \cdot & \cdot & \cdot \\ \hline
    \cdot      & \cdot & \cdot & \cdot \\
    \pm 1 & \cdot & \cdot & 1 \end{array} \right)\ ,
\end{equation}
and for $|\varphi^\pm\> = (|01\> \pm |10\>)/\sqrt{2}$
\begin{equation}
|\varphi^\pm\>\<\varphi^\pm| = \frac 12 \left(
\begin{array}{cc|cc}  \cdot & \cdot & \cdot & \cdot \\
    \cdot      & 1 & \pm 1 & \cdot \\ \hline
    \cdot      & \pm 1 & 1 & \cdot \\
    \cdot & \cdot & \cdot & \cdot \end{array} \right)\ .
\end{equation}

2. Werner state \cite{Werner}
\begin{equation}\label{Werner-2}
\mathcal{W} = \frac 14 \left( \begin{array}{cc|cc}
    1-p & \cdot & \cdot & \cdot \\
    \cdot      & 1+p & -2p & \cdot \\ \hline
    \cdot      & -2p & 1+p & \cdot \\
    \cdot & \cdot & \cdot & 1-p \end{array} \right)\ ,
\end{equation}
with $-1/3 \leq p \leq 1$. PPT condition implies well known result
$p \leq 1/3$.

3. Isotropic state  \cite{ISO}
\begin{equation}\label{}
\mathcal{I} = \frac 14 \left( \begin{array}{cc|cc}
    1+p & \cdot & \cdot & 2p \\
    \cdot      & 1-p & \cdot & \cdot \\ \hline
    \cdot      & \cdot & 1-p & \cdot \\
    2p & \cdot & \cdot & 1+p \end{array} \right)\ ,
\end{equation}
with $-1/3 \leq p \leq 1$. Again PPT condition implies well known
result $p \leq 1/3$.

4. $O(2)\ot O(2)$--invariant state
\begin{equation}\label{}
\mathcal{O} = \frac 14 \left( \begin{array}{cc|cc}
    a +2b & \cdot & \cdot & 2b-a \\
    \cdot      & a+2c & a-2c & \cdot \\ \hline
    \cdot      & a-2c & a+2c & \cdot \\
    2b-a & \cdot & \cdot & a+2b \end{array} \right)\ ,
\end{equation}
with $a,b,c\geq 0$ and $a+b+c=1$. It is clear that $\mathcal{O}$
is positive  and $\mathcal{O}$ is PPT iff
\begin{equation}\label{}
    b \leq \frac 12\ , \ \ \ c \leq \frac 12\ ,
\end{equation}
which reproduces well known result \cite{Werner2}.

\subsection{3 qubits}

There are in principle two ways to generalize the 2-qubit
circulant decomposition for the case of three qubits. Either one
decomposes $\mathcal{H}_{2^3} = \mathbb{C}^2 \ot \mathbb{C}^2 \ot
\mathbb{C}^2$ into four 2-dimensional subspaces or into two
4-dimensional ones.

\subsubsection{``$\ 8 = 2 \oplus 2 \oplus 2 \oplus 2$''}

Let us define 2-dimensional subspace
\begin{equation}\label{}
 \Delta_{00} =
    \mbox{span}\left\{ e_0 \ot e_0 \ot e_0\, , e_1 \ot e_1 \ot e_1 \right\}
\end{equation}
and for any two binaries $\mu$ and $\nu$ define
\begin{equation}\label{}
  \Delta_{\mu\nu} = (\mathbb{I} \ot S^\mu \ot S^\nu)\Delta_{00}\ .
\end{equation}
One easily finds
\begin{eqnarray}\label{}
     \Delta_{01} &=&
    \mbox{span}\left\{ e_0 \ot e_0 \ot e_1\, , e_1 \ot e_1 \ot e_0 \right\} \
    , \nonumber    \\
     \Delta_{10} &=&
    \mbox{span}\left\{ e_0 \ot e_1 \ot e_0\, , e_1 \ot e_0 \ot e_1 \right\} \
    ,     \\
     \Delta_{11} &=&
    \mbox{span}\left\{ e_0 \ot e_1 \ot e_1\, , e_1 \ot e_0 \ot e_0 \right\} \
    . \nonumber
\end{eqnarray}
It is clear that
\begin{equation}\label{}
\mathcal{H}_{2^3} = \Delta_{00} \oplus \Delta_{01} \oplus
\Delta_{10} \oplus \Delta_{11} \ .
\end{equation}
Now, we construct 3-qubit density matrix $\rho$ of the following
form
\begin{equation}\label{}
\rho = \rho_{00} + \rho_{01} + \rho_{10} + \rho_{11} \ ,
\end{equation}
where each $\rho_{\mu\nu}$ is supported on $\Delta_{\mu\nu}$. One
has therefore
\begin{equation}\label{}
    \rho_{\mu\nu} = \sum_{i,j=0}^1 \, x^{(\mu\nu)}_{ij} \, e_{ij}
    \ot S^\mu e_{ij} S^{\mu *} \ot S^\nu e_{ij} S^{\nu *} \ ,
\end{equation}
which generalizes 2-qubit construction (\ref{ro-2-qubit}).
Positivity of $\rho$ is guarantied by positivity of each $2\times
2$ matrix $[x^{(\mu\nu)}]$, and the normalization
$\mbox{Tr}\rho=1$ is equivalent to
\begin{equation}\label{}
    \mbox{Tr}\left( x^{(00)} + x^{(01)} + x^{(01)} + x^{(11)} \right) = 1\
    .
\end{equation}
 One obtains therefore the following
block matrix
\begin{widetext}
\begin{equation}\label{3qubit-2222}
 \hspace*{-.1cm}
  \rho \ =\  \left( \begin{array}{cc|cc||cc|cc}
    x^{(00)}_{00} & \cdot & \cdot & \cdot & \cdot & \cdot & \cdot &  x^{(00)}_{01} \\
    \cdot& x^{(01)}_{00} & \cdot & \cdot & \cdot & \cdot & x^{(01)}_{01}  &
    \cdot \\ \hline
    \cdot & \cdot & x^{(10)}_{00} & \cdot & \cdot & x^{(10)}_{01} & \cdot &  \cdot \\
    \cdot & \cdot & \cdot & x^{(11)}_{00} & x^{(11)}_{01} & \cdot & \cdot &
    \cdot \\ \hline  \hline
    \cdot & \cdot & \cdot & x^{(11)}_{10} & x^{(11)}_{11} & \cdot & \cdot &
    \cdot \\
    \cdot & \cdot & x^{(10)}_{10} & \cdot & \cdot & x^{(10)}_{11} & \cdot & \cdot
    \\ \hline
    \cdot & x^{(01)}_{10} & \cdot & \cdot & \cdot & \cdot & x^{(01)}_{11}  &
    \cdot \\
    x^{(00)}_{10} & \cdot & \cdot & \cdot & \cdot & \cdot & \cdot &  x^{(00)}_{11}
    \end{array} \right)\ ,
\end{equation}
\end{widetext} where vertical and horizontal lines remind us about
the splitting into blocks corresponding to tensor product
structure $\mathbb{C}^2 \ot \mathbb{C}^2 \ot \mathbb{C}^2$. Double
lines introduce splitting into four blocks corresponding to
$\mathbb{C}^2 \ot (\mathbb{C}^2 \ot \mathbb{C}^2)$ and then single
lines introduce splitting within each $4 \times 4$  block
corresponding to the second tensor product $(\mathbb{C}^2 \ot
\mathbb{C}^2)$.  Now, let us perform partial transposition. There
are three independent transformations
\begin{equation}\label{}
    \tau_{\alpha\beta} = \oper \ot \tau^\alpha \ot \tau^\beta\ ,
\end{equation}
with binary indices $\alpha$ and $\beta$:
\begin{eqnarray}\label{}
 \tau_{01} = \oper \ot \oper \ot \tau\ , \nonumber \\
 \tau_{10} = \oper \ot \tau \ot \oper\ , \\
 \tau_{11} = \oper \ot \tau \ot \tau\ . \nonumber
\end{eqnarray}
Now, it is easy to see that  $ \tau_{\alpha\beta}\rho$ has the
same circular structure as original $\rho$ defined in (\ref{3C})
with new $2\times 2$ matrices $y^{(\mu\nu)[\alpha\beta]}$, that
is,
\begin{equation}\label{}
  \tau_{\alpha\beta}\, \rho = \sum_{\mu,\nu=0}^1 \sum_{i,j=0}^1 \, y^{(\mu\nu)[\alpha\beta]}_{ij} \, e_{ij}
    \ot S^\mu e_{ij} S^{\mu *} \ot S^\nu e_{ij} S^{\nu *} \ .
\end{equation}

\begin{proposition}
The set of $2 \times 2$ matrices $y^{(\mu\nu)[\alpha\beta]}$ is
given by
\begin{equation}\label{}
y^{(\mu\nu)[\alpha\beta]} = x^{(\mu\nu)} \circ \mathbb{I} +
x^{(\mu+\alpha,\nu+\beta)} \circ S \ ,
\end{equation}
with addition modulo $2$.
\end{proposition}
Using straightforward definition that $\rho$ is
$(\alpha\beta)$-PPT if $\tau_{\alpha\beta}\,\rho \geq 0$, one has
the following

\begin{theorem}
A 3-qubit circulant state $\rho$ is $(\alpha\beta)$-{\rm PPT} iff
\[ y^{(\mu\nu)[\alpha\beta]} \geq 0 \ ,  \]
for all binary $\mu$ and $\nu$.
\end{theorem}

It is clear that (\ref{3qubit-2222}) generalizes 2-qubit circulant
state (\ref{2C-T}). Note, however, that reducing 3-qubit state
with respect to one subsystem one ends up with the following
separable 2-qubit state
\begin{equation}\label{}
\mbox{Tr}_1 \rho = \sum_{\mu,\nu=0}^1 \sum_{i,j=0}^1
x^{(\mu\nu)}_{ii} e_{ii} \ot e_{ii} \ .
\end{equation}
It is again circulant but has very special structure: $[a_{ij}]$
is diagonal with
\[ a_{ii} = \sum_{\mu,\nu=0}^1
x^{(\mu\nu)}_{ii}\ , \] and $[b_{ij}]=0$. It is therefore clear
that a general 2-qubit circulant state can not be obtained via
reduction of (\ref{3qubit-2222}).

{\it Examples.} 1. GHZ state \cite{GHZ}

\begin{equation}\label{}
    |\rm{GHZ}\> = \frac{1}{\sqrt{2}} (|000\> + |111\> )\ ,
\end{equation}
does belong to circulant class which is easily seen from the
corresponding density matrix
\begin{equation}\label{}
    x^{(\mu\nu)} =\, \delta_{\mu 0}\, \delta_{\nu 0}\, \left( \begin{array}{cc} 1 & 1 \\ 1 & 1 \end{array}
    \right) \ .
\end{equation}

On the other hand the well known W state
\begin{equation}\label{}
    |\rm{W}\> = \frac{1}{\sqrt{3}} (|001\> + |010\> + |100\> )\ ,
\end{equation}
is not circulant. The corresponding density matrix reads as
follows
\begin{equation}\label{W-state}
 \hspace*{-.1cm}
  \rho_{\rm W} \ =\  \frac 13 \left( \begin{array}{cc|cc||cc|cc}
    \cdot & \cdot & \cdot & \cdot & \cdot & \cdot & \cdot &  \cdot \\
    \cdot & 1 & 1 & \cdot & 1 & \cdot & \cdot  &
    \cdot \\ \hline
    \cdot & 1 & 1 & \cdot & 1 & \cdot & \cdot  &
    \cdot \\
    \cdot& \cdot & \cdot & \cdot & \cdot & \cdot & \cdot  &
    \cdot \\ \hline\hline
    \cdot& 1 & 1 & \cdot & 1 & \cdot & \cdot  &
    \cdot \\
    \cdot& \cdot & \cdot & \cdot & \cdot & \cdot & \cdot  &
    \cdot \\ \hline
    \cdot & \cdot & \cdot & \cdot & \cdot & \cdot & \cdot  & \cdot
    \\
    \cdot& \cdot & \cdot & \cdot & \cdot & \cdot & \cdot  &
    \cdot
    \end{array} \right)\ .
\end{equation}

2. Bell states: the following 8 vectors
\begin{equation}\label{}
    \psi_{\alpha\beta\gamma} = (-1)^\alpha\, ( \mathbb{I} \ot S^\beta
    \ot S^\gamma) |\rm{GHZ}\> \ ,
\end{equation}
with binary $\alpha,\beta,\gamma$ define circulant states. These
are 3-qubit generalization of 2-qubit Bell states. Note that
\begin{equation}\label{}
 \psi_{\alpha\beta\gamma} \in \Delta_{\beta\gamma}\ ,
\end{equation}
and the corresponding matrices $x^{(\mu\nu)}$ read as follows
\begin{equation}\label{}
    x^{(\mu\nu)} =\, \delta_{\mu\beta }\, \delta_{\nu\gamma }\, \left( \begin{array}{cc} 1 & (-1)^\alpha\\ (-1)^\alpha & 1 \end{array}
    \right) \ .
\end{equation}

 3. Generalized 3-qubit isotropic state

\begin{equation}\label{}
    \rho = \frac{1-s}{2^3}\,\mathbb{I}^{\ot 3} + s \,|\rm{GHZ}\>\<\rm{GHZ}|
\end{equation}
with $s\in [-1/7,1]$. One finds for $x^{(\mu\nu)}$ matrices
\begin{equation}\label{}
    x^{(00)} = \frac 18 \left( \begin{array}{cc} 1+3s & 4s \\ 4s &
    1+3s \end{array} \right)\ ,
\end{equation}
and
\begin{equation}\label{}
    x^{(01)} =   x^{(10)} =  x^{(11)} = \frac{1-s}{8}\,\mathbb{I}\
    .
\end{equation}
The only nontrivial PPT condition comes from the positivity of
\begin{equation}\label{}
    y^{(01)[01]} = \frac 18 \left( \begin{array}{cc} 1-s & 4s \\ 4s & 1-s  \end{array} \right)\ ,
\end{equation}
which implies $s \leq 1/5$. Actually, it is well known \cite{Art}
that $\rho$ is fully separable iff $s \leq 1/5$.

4. 2-parameter 3-qubit state from \cite{Art2}:

\begin{equation}\label{Art-3-qubit}
 \hspace*{-.1cm}
  \rho(c,d) \ =\  \frac 18 \left( \begin{array}{cc|cc||cc|cc}
    1 & \cdot & \cdot & \cdot & \cdot & \cdot & \cdot &  1 \\
    \cdot & 1 & \cdot & \cdot & \cdot & \cdot  & 1 &
    \cdot \\ \hline
    \cdot & \cdot & 1 & \cdot & \cdot & c & \cdot  &
    \cdot \\
    \cdot & \cdot & \cdot & 1 & d & \cdot & \cdot  &
    \cdot \\ \hline\hline
    \cdot & \cdot & \cdot & d & 1 & \cdot & \cdot  &
    \cdot \\
    \cdot & \cdot & c & \cdot & \cdot & 1 & \cdot  &
    \cdot \\ \hline
    \cdot & 1 & \cdot & \cdot & \cdot & \cdot & 1  & \cdot
    \\
    1 & \cdot & \cdot & \cdot & \cdot & \cdot & \cdot  &
    1
    \end{array} \right)\ ,
\end{equation}
with $-1/8 \leq c,d \leq 1/8$.  It was shown \cite{Art2} that
$\rho(c,d)$ has positive partial transposes iff $c=d$. Moreover,
this condition implies full separability.

\subsubsection{``$\ 8 = 4 \oplus 4$''}

There are two ways to construct 4-dimensional circulant
decompositions of 3-qubit Hilbert space out of 2-dimensional
mutually orthogonal spaces $\Delta_{\mu\nu}$: either one
introduces
\begin{eqnarray}\label{}
    \Sigma_0 &=& \Delta_{00} \oplus \Delta_{11}\ , \\
    \Sigma_1 &=& \Delta_{01} \oplus \Delta_{10}\ ,
\end{eqnarray}
or
\begin{eqnarray}\label{}
    \Xi_0 &=& \Delta_{00} \oplus \Delta_{10} \ , \\
    \Xi_1 &=& \Delta_{01} \oplus \Delta_{11} \ .
\end{eqnarray}
The construction is clear:
\begin{eqnarray}\label{}
    \Sigma_0 &=& \bigoplus_{\mu+\nu=0} \Delta_{\mu\nu} \ , \\
    \Sigma_1 &=& \bigoplus_{\mu+\nu=1} \Delta_{\mu\nu} ,
\end{eqnarray}
whereas a second decomposition uses the following scheme
\begin{eqnarray}\label{}
    \Xi_\nu &=& \bigoplus_{\mu=0}^1 \Delta_{\mu\nu} \ .
\end{eqnarray}
Note, that using binary codes,  $\Xi_0$ is constructed out of
$\Delta_{\mu\nu}$ with $\mu\nu$ representing binary code for `$0$'
(mod 2) and $\Xi_1$ is constructed out of $\Delta_{\mu\nu}$ with
$\mu\nu$ representing binary code for `$1$' (mod 2). One easily
finds
\begin{widetext}
\begin{eqnarray}\label{}
\Sigma_0 &=& \mbox{span}\left\{ e_0 \ot e_0 \ot e_0\, ,e_1 \ot e_1
\ot e_1\, , e_1 \ot e_0 \ot e_0\, , e_0 \ot e_1 \ot e_1 \right\} \
    , \nonumber    \\
    \Sigma_1 &=& \mbox{span}\left\{ e_0 \ot e_0 \ot e_1\, ,e_1 \ot
    e_1 \ot e_0\, , e_1 \ot e_0 \ot e_1\, , e_0 \ot e_1 \ot e_0 \right\} \
    ,
\end{eqnarray}
\end{widetext}
and
\begin{widetext}
\begin{eqnarray}\label{}
\Xi_0 &=& \mbox{span}\left\{ e_0 \ot e_0 \ot e_0\, ,e_1 \ot e_1
\ot e_1\, , e_0 \ot e_1 \ot e_0\, , e_1 \ot e_0 \ot e_1 \right\} \
    , \nonumber    \\
    \Xi_1 &=& \mbox{span}\left\{ e_0 \ot e_0 \ot e_1\, ,e_1 \ot
    e_1 \ot e_0\, , e_0 \ot e_1 \ot e_1\, , e_1 \ot e_0 \ot e_0 \right\} \
    .
\end{eqnarray}
\end{widetext}

It is clear that both decompositions  are circulant, that is,
\begin{equation}\label{}
    \Sigma_1 = (\mathbb{I} \ot \mathbb{I}
\ot S) \Sigma_0\ ,
\end{equation}
and
\begin{equation}\label{}
    \Xi_1 = (\mathbb{I} \ot \mathbb{I}
\ot S) \Xi_0\ .
\end{equation}
Now we are ready to construct  the corresponding 3-qubit circulant
states: a circulant state corresponding to
\[ \Sigma_0 \oplus \Sigma_1 = \mathcal{H}_{2^3} \ , \]
is defined by
\[  \sigma = \sigma_0 + \sigma_1\ , \]
with $\sigma_\alpha$ supported on $\Sigma_\alpha$. To define
$\rho_0$ and $\rho_1$ one needs $4 \times 4$ matrices $a$ and $b$.
 One has
\begin{eqnarray}\label{rho0-3}
\sigma_0 &=& \sum_{i,j=0}^1\, \sum_{\mu,\nu=0}^1\, a_{\mu i;\nu
j}\, e_{\mu\nu} \ot e_{ij} \ot e_{ij}\ , \\ \label{rho1-3}
\sigma_1 &=& \sum_{i,j=0}^1\, \sum_{\mu,\nu=0}^1\, b_{\mu i;\nu
j}\, e_{\mu\nu} \ot e_{ij} \ot S\,e_{ij} S^*\ ,
\end{eqnarray}
where $a_{\mu i;\nu j}$ and $b_{\mu i;\nu j}$ are matrix elements
of $a$ and $b$ considered as matrices in the tensor product $M_4 =
M_2 \ot M_2 =  M_2(M_2)$.  One obtains therefore the following
block matrix
\begin{widetext}
\begin{equation}\label{3qubit-44}
 \hspace*{-.1cm}
  \sigma \ =\  \left( \begin{array}{cc|cc||cc|cc}
    a_{00;00} & \cdot &  \cdot & a_{00;01} & a_{00;10} & \cdot &  \cdot &  a_{00;11} \\
    \cdot& b_{00;00} & b_{00;01} & \cdot & \cdot & b_{00;10} & b_{00;11}  &
    \cdot \\ \hline
    \cdot & b_{01;00} & b_{01;01} & \cdot & \cdot & b_{01;10} &  b_{01;11} & \cdot  \\
    a_{01;00} & \cdot &  \cdot & a_{01;01} & a_{01;10} & \cdot & \cdot & a_{01;11}  \\ \hline  \hline
    a_{10;00} & \cdot &  \cdot & a_{10;01} & a_{10;10} & \cdot &  \cdot &  a_{10;11} \\
    \cdot& b_{10;00} & b_{10;01} & \cdot & \cdot & b_{10;10} & b_{10;11}  &
    \cdot \\ \hline
    \cdot & b_{11;00} & b_{11;01} & \cdot & \cdot & b_{11;10} &  b_{11;11} & \cdot  \\
    a_{11;00} & \cdot &  \cdot & a_{11;01} & a_{11;10} & \cdot & \cdot & a_{11;11}
    \end{array} \right)\ .
\end{equation}
\end{widetext}
Let us note, that the density matrix defined by
(\ref{3qubit-2222}) is a special case of (\ref{3qubit-44}) where
the matrices $a$ and $b$ are given by:
\begin{equation}\label{}
    a = \left( \begin{array}{cc|cc}
     x^{(00)}_{00} & \cdot & \cdot & x^{(00)}_{01} \\
     \cdot & x^{(11)}_{00} & x^{(11)}_{01} & \cdot \\ \hline
     \cdot & x^{(11)}_{10} & x^{(11)}_{11} & \cdot \\
     x^{(00)}_{10} & \cdot & \cdot & x^{(00)}_{11} \end{array} \right)\ ,
\end{equation}
and
\begin{equation}\label{}
    b = \left( \begin{array}{cc|cc}
      x^{(01)}_{00} & \cdot & \cdot & x^{(01)}_{01} \\
     \cdot & x^{(10)}_{00} & x^{(10)}_{01} & \cdot \\ \hline
     \cdot & x^{(10)}_{10} & x^{(10)}_{11} & \cdot \\
     x^{(01)}_{10} & \cdot & \cdot & x^{(01)}_{11} \end{array}
     \right)\
     .
\end{equation}
Similarly, a 3-qubit circulant state corresponding to
\[ \Xi_0 \oplus \Xi_1 = \mathcal{H}_{2^3} \ , \]
is defined by
\[  \xi = \xi_0 + \xi_1\ , \]
with $\xi_\alpha$ supported on $\Xi_\alpha$. It is defined via two
$4 \times 4$ matrices $c$ and $d$:
\begin{eqnarray}\label{}
\xi_0 &=& \sum_{i,j=0}^1\, \sum_{\mu,\nu=0}^1\, c_{i\mu ;j\nu }\,
 e_{ij} \ot e_{\mu\nu} \ot e_{ij}\ ,\nonumber  \\ \label{xi-3} \xi_1 &=&
\sum_{i,j=0}^1\, \sum_{\mu,\nu=0}^1\, d_{i\mu ;j\nu }\,  e_{ij}
\ot e_{\mu\nu} \ot S\,e_{ij} S^*\ .
\end{eqnarray}
One finds
\begin{widetext}
\begin{equation}\label{3C}
 \hspace*{-.1cm}
 \xi \ =\  \left( \begin{array}{cc|cc||cc|cc}
    c_{00;00} & \cdot & c_{00;01} & \cdot & \cdot & c_{00;10} & \cdot &  c_{00;11} \\
    \cdot& d_{00;00} & \cdot & d_{00;01} & d_{00;10} & \cdot & d_{00;11}  &
    \cdot \\ \hline
    c_{01;00} & \cdot & c_{01;01} & \cdot & \cdot & c_{01;10} & \cdot &  c_{01;11} \\
    \cdot & d_{01;00} & \cdot & d_{01;01} & d_{01;10} & \cdot & d_{01;11}  &
    \cdot \\ \hline  \hline
    \cdot & d_{10;00} & \cdot & d_{10;01} & d_{10;10} & \cdot & d_{10;11}  &
    \cdot \\
    c_{10;00} & \cdot & c_{10;01} & \cdot & \cdot & c_{10;10} & \cdot &  c_{10;11}
    \\ \hline
    \cdot & d_{11;00} & \cdot & d_{11;01} & d_{11;10} & \cdot & d_{11;11}  &
    \cdot \\
    c_{11;00} & \cdot & c_{11;01} & \cdot & \cdot & c_{11;10} & \cdot &  c_{11;11}
    \end{array} \right)\ .
\end{equation}
\end{widetext}
It is clear that  partially transposed $\tau_{\alpha\beta} \sigma$
and $\tau_{\alpha\beta}\xi$ have exactly the same circulant
structure as original $\sigma$ and $\xi$. One easily finds for the
corresponding partial transpositions:
\begin{eqnarray}\label{}
\tau_{\alpha\beta}\,\xi &=& \sum_{i,j=0}^1\, \sum_{\mu,\nu=0}^1\,
c_{i\mu ;j\nu }^{[\alpha\beta]}\,  e_{ij} \ot e_{\mu\nu} \ot e_{ij} \nonumber \\
&+& \sum_{i,j=0}^1\, \sum_{\mu,\nu=0}^1\, d_{i\mu ;j\nu
}^{[\alpha\beta]}\,  e_{ij} \ot e_{\mu\nu} \ot S\,e_{ij} S^*\ ,
\end{eqnarray}
where the new $4 \times 4$ matrices $c_{\mu i;\nu
j}^{[\alpha\beta]}$ and $d_{\mu i;\nu j}^{[\alpha\beta]}$ are
given by:
\begin{equation}\label{}
    c^{[00]} = c \ , \ \ \ \  d^{[00]} = d\ ,
\end{equation}
and
\begin{eqnarray}\label{}
    c^{[10]} = (\oper \ot \tau)\,c\ , \ \ \ \
    d^{[10]} = (\oper \ot \tau)\, d\ ,
\end{eqnarray}
where
\begin{equation*}\label{}
[(\oper \ot \tau)\, c]_{i\mu ;j\nu } = c_{i\nu ;j\mu }\ , \ \ \
[(\oper \ot \tau)\, d]_{i\mu ;j\nu } = d_{i\nu ;j\mu }\ .
\end{equation*}
 Moreover,
\begin{eqnarray}\label{}
 c^{[\alpha 1]} &=& c^{[\alpha 0]} \circ (\mathbb{I} \ot \widetilde{\mathbb{I}}) + d^{[\alpha 0]} \circ (S
 \ot \widetilde{\mathbb{I}})\ ,\nonumber \\
d^{[\alpha 1]} &=& d^{[\alpha 0]} \circ (\mathbb{I} \ot
\widetilde{\mathbb{I}}) + c^{[\alpha 0]} \circ (S \ot
\widetilde{\mathbb{I}})\ ,
\end{eqnarray}
where
\begin{equation}\label{tilde-I-2}
\widetilde{\mathbb{I}} = \mathbb{I} + S\ =\ \left(
\begin{array}{cc} 1 & 1
\\ 1 & 1
\end{array} \right)\  .
\end{equation}
The similar formulae one easily finds for $a^{[\alpha\beta]}$ and
$b^{[\alpha\beta]}$ defined by
\begin{eqnarray}\label{}
\tau_{\alpha\beta}\,\sigma &=& \sum_{i,j=0}^1\,
\sum_{\mu,\nu=0}^1\,
a_{\mu i;\nu j}^{[\alpha\beta]}\,  e_{\mu\nu} \ot e_{ij} \ot  e_{ij} \nonumber \\
&+& \sum_{i,j=0}^1\, \sum_{\mu,\nu=0}^1\, b_{\mu i;\nu
j}^{[\alpha\beta]}\,  e_{\mu\nu} \ot e_{ij}  \ot S\,e_{ij} S^*\ .
\end{eqnarray}
\begin{theorem}
A circulant 3-qubit state $\sigma$ is $(\alpha\beta)$--{\rm PPT}
iff $a^{[\alpha\beta]}$ and $b^{[\alpha\beta]}$ are semi-positive
matrices. Similarly, a circulant 3-qubit state $\xi$ is
$(\alpha\beta)$--{\rm PPT} iff $c^{[\alpha\beta]}$ and
$d^{[\alpha\beta]}$ are semi-positive matrices.
\end{theorem}

Let us observe that a 3-qubit circulant state may be reduced to
the 2-qubit circulant state. Consider for example a density
operator $\xi$ defined in (\ref{xi-3}). Note that reduction with
respect to the second factor gives
\begin{eqnarray}\label{}
\mbox{Tr}_2\xi_0 &=& \sum_{i,j=0}^1\, \sum_{\mu=0}^1\,
c_{i\mu;j\mu}\, e_{ij} \ot e_{ij} \nonumber \\ &=& \sum_{i,j=0}^1\, ({\rm{Tr}}_2 c)_{ij}\, e_{ij} \ot e_{ij}\ , \\
 \mbox{Tr}_2\xi_1 &=& \sum_{i,j=0}^1\, \sum_{\mu=0}^1\,
d_{i\mu;j\mu}\, e_{ij}  \ot S\,e_{ij} S^* \nonumber \\ &=&
\sum_{i,j=0}^1\, ({\rm{Tr}}_2 d)_{ij}\, e_{ij} \ot S\,e_{ij} S^*\
,
\end{eqnarray}
and hence
\begin{equation}\label{}
\mbox{Tr}_2\xi = \mbox{Tr}_2\xi_0 + \mbox{Tr}_2\xi_1 \ ,
\end{equation}
is a 2-qubit circulant state. It is no longer true for the
remaining reductions with respect to the first and third factors.
One obtains
\begin{eqnarray}\label{}
\mbox{Tr}_1\xi = \sum_{i=0}^1  \sum_{\mu,\nu=0}^1\, [c
+d]_{i\mu;i\nu}\, e_{\mu\nu} \ot e_{ii}  \ ,
\end{eqnarray}
and
\begin{eqnarray}\label{}
\mbox{Tr}_3\xi = \sum_{i=0}^1 \sum_{\mu,\nu=0}^1\,
[c+d]_{i\mu;i\nu}\, e_{ii} \ot e_{\mu\nu}    \ ,
\end{eqnarray}
which are not circulant  states.

\subsection{$N$ qubits}

Consider now a general case of $N$ qubits living in
$(\mathbb{C}^d)^{\ot N}$. Again, there are two natural ways to
decompose the corresponding Hilbert space $\mathcal{H}_{2^N}$:
either into $2^{N-1}$ two-dimensional subspaces or into two
$2^{N-1}$--dimensional subspaces.

\subsubsection{``$\ 2^N = 2 \oplus 2 \oplus \ldots  \oplus 2$''}

Let us introduce a circulant decomposition of $\mathcal{H}_{2^N}$
into $2^{N-1}$ two-dimensional subspaces. Now each integer from
the set $\{0,1,\ldots,2^{N-1}-1\}$ may be represented by a string
of $N-1$ binaries $(\mu_1\ldots\mu_{N-1})$. Let us define
2-dimensional subspace
\begin{equation}\label{}
 \Delta_{0\ldots 0} =
    \mbox{span}\left\{ e_0 \ot \ldots \ot e_0\, , e_1 \ot \ldots \ot e_1
    \right\}\ ,
\end{equation}
and for any string of binaries $(\mu_1\ldots\mu_{N-1})$ define
\begin{equation}\label{}
  \Delta_{\mu_1\ldots\mu_{N-1}} = (\mathbb{I} \ot S^{\mu_1} \ot \ldots  \ot S^{\mu_{N-1}})\Delta_{0\ldots 0}\ .
\end{equation}
Introducing convenient vector notation
\[ \bmu = (\mu_1,\ldots,\mu_{N-1})\ , \]
one has
\begin{equation}\label{}
    \Delta_\sbmu = (\oper \ot S^\sbmu)\Delta_{\bf 0}\ ,
\end{equation}
with
\[ S^\sbmu = S^{\mu_1} \ot \ldots  \ot S^{\mu_{N-1}}\ , \]
and $\Delta_{\bf 0} = \Delta_{0\ldots 0}$. One clearly has
\begin{equation}\label{N-DS}
    \mathcal{H}_{2^N} = \bigoplus_\sbmu\, \Delta_\sbmu\ ,
\end{equation}
where the sum runs over all binary $(N-1)$--vectors $\bmu$.

Now, let us construct a circulant $N$-qubit state $\rho$ based on
(\ref{N-DS}):
\begin{equation}\label{}
    \rho = \sum_\sbmu\, \rho_\sbmu\ ,
\end{equation}
where each $\rho_\sbmu$ is supported on $\Delta_\sbmu$. One has
therefore
\begin{equation}\label{}
    \rho_\sbmu = (\oper \ot S^\sbmu) \left[ \sum_{i,j=0}^1\, x^{(\sbmu)}_{ij} \ e_{ij} \ot
    \ldots \ot e_{ij} \right](\oper \ot S^\sbmu)^* \ ,
\end{equation}
where $[x^{(\sbmu)}]$ are $2\times 2$ semi-positive matrices.
Normalization of $\rho$ implies
\begin{equation}\label{}
     \sum_\sbmu\, \mbox{Tr}\,x^{(\sbmu)} = 1\ .
\end{equation}
Now, partial transpositions are labeled by a binary
$(N-1)$--vectors $\bsigma=(\sigma_1,\ldots,\sigma_{N-1})$
\begin{equation}\label{multi-PP}
    \tau_\sbsigma = \oper \ot \tau^{\sigma_1} \ot \ldots \ot
    \tau^{\sigma_{N-1}}\ .
\end{equation}
Note, that each partial transposition $\tau_\sbsigma \rho$ belongs
to the same class of circulant states
\begin{equation}\label{}
    \tau_\sbsigma \rho = \sum_\sbmu\, \rho_\sbmu^{[\sbsigma]}\ ,
\end{equation}
with
\begin{equation}\label{}
    \rho_\sbmu^{[\sbsigma]} = (\oper \ot S^\sbmu) \left[ \sum_{i,j=0}^1\, y^{(\sbmu)[\sbsigma]}_{ij} \ e_{ij} \ot
    \ldots \ot e_{ij} \right](\oper \ot S^\sbmu)^* \ ,
\end{equation}
where the new $2 \times 2$ matrices $y^{(\sbmu)[\sbsigma]}$ are
given by the following formula
\begin{equation}\label{x-mu-2}
    y^{(\sbmu)[\sbsigma]} = x^{[\sbmu]} \circ \mathbb{I} +  x^{[\sbmu + \sbsigma]}
    \circ S\ .
\end{equation}
 A state $\rho$ is $\!\bsigma$--PPT iff $\tau_\sbsigma
\rho \geq 0$ and hence one has
\begin{theorem}
A circulant state $\rho$ is $\!\bsigma$-{\rm PPT} iff
$y^{(\sbmu)[\sbsigma]}$ are semi-positive for all $\bmu$.
\end{theorem}

{\it Examples.} 1. Generalized GHZ state \cite{GHZ}

\begin{equation}\label{}
    |\rm{GHZ}\> = \frac{1}{\sqrt{2}} (|0\ldots 0\> + |1\ldots 1\> )\ ,
\end{equation}
does belong to circulant class which is easily seen from the
corresponding density matrix
\begin{equation}\label{}
    x^{(\sbmu)} =\, \delta{\sbmu,\sbnu}\, \widetilde{\mathbb{I}} \ .
\end{equation}

2. Generalized Bell states: the following $2^N$ vectors
\begin{equation}\label{2N-Bell}
    \psi_{\alpha\sbnu} =  (-1)^\alpha( \mathbb{I} \ot S^\sbnu) |\rm{GHZ}\> \ ,
\end{equation}
with $\alpha=0,1$ and  binary $(N-1)$--vector $\!\bnu$ define
circulant states. These are $N$-qubit generalization of 2-qubit
Bell states. Note that
\begin{equation}\label{}
 \psi_{\alpha\sbnu} \in \Delta_{\sbnu}\ ,
\end{equation}
and the corresponding matrices $x^{(\sbmu)}$ read as follows
\begin{equation}\label{2x2}
    x^{(\sbmu)} =\, \delta_{\sbmu,\sbnu}\,
    \left( \begin{array}{cc} 1 & (-1)^\alpha\\ (-1)^\alpha & 1 \end{array}
    \right) \ .
\end{equation}

 3. Generalized $N$-qubit isotropic state

\begin{equation}\label{}
    \rho = \frac{1-s}{2^N}\,\mathbb{I}^{\ot N} + s \,|\rm{GHZ}\>\<\rm{GHZ}|
\end{equation}
with $s\in [-1/(2^N-1),1]$. One finds for $x^{(\sbmu)}$ matrices
\begin{equation}\label{}
    x^{({\bf 0})} = \frac{1}{2^N} \left( \begin{array}{cc} 1+(2^N-1)s & 2^{N-1}s \\ 2^{N-1}s &
    1+(2^N-1)s \end{array} \right)\ ,
\end{equation}
and
\begin{equation}\label{}
    x^{(\sbmu)} =  \frac{1-s}{2^N}\,\mathbb{I}\
    ,
\end{equation}
for $\bmu \neq {\bf 0}$. The only nontrivial PPT condition comes
from the positivity of
\begin{equation}\label{}
 \left( \begin{array}{cc} 1-s & 2^{N-1}s \\ 2^{N-1}s & 1-s  \end{array} \right)\ ,
\end{equation}
which implies
\[ s \leq \frac{1}{2^{N-1} +1}\ .\]
The above condition guaranties full $N$-separability of $\rho$
\cite{Art}.

4. 2-parameter $N$-qubit state from \cite{Art3}: for $-1/2^N\leq
c,d \leq 1/2^N$ one defines a set of matrices $x^{(\sbmu)}$

\begin{equation}\label{}
 x^{(\sbmu)} = \frac{1}{2^N} \left( \begin{array}{cc} 1 & 1 \\ 1 &
    1 \end{array} \right)\ ,
\end{equation}
for $\!\bmu$ corresponding to binary representation of
$\{0,1,\ldots,2^{N-2}-1\}$,
\begin{equation}\label{}
 x^{(\sbmu)} = \frac{1}{2^N} \left( \begin{array}{cc} 1 & c \\ c &
    1 \end{array} \right)\ ,
\end{equation}
for $\!\bmu$ corresponding to binary representation of
$\{2^{N-2}-1,\ldots,2^{N-1}-1\}$, and
\begin{equation}\label{}
 x^{(\sbmu)} = \frac{1}{2^N} \left( \begin{array}{cc} 1 & d \\ d &
    1 \end{array} \right)\ ,
\end{equation}
for $\!\bmu$ corresponding to binary representation of
$\{2^{N-1}-1,\ldots,2^{N}-1\}$. It generalizes a 3-qubit state
defined in (\ref{Art-3-qubit}). It was shown \cite{Art3} that the
above $N$-qubit circulant state has positive partial transposes
iff $c=d$. Moreover, this condition implies full separability.

\subsubsection{``$\ 2^N = 2^{N-1} \oplus 2^{N-1}$''}

There are several ways to construct $2^{N-1}$--dimensional
circulant decompositions of $N$-qubit Hilbert space out of
2-dimensional mutually orthogonal spaces $\Delta_{\sbmu}$. The
following choice
\begin{eqnarray}\label{}
    \Sigma_0 &=& \bigoplus_{|\sbmu|=0} \Delta_{\sbmu} \ , \\
    \Sigma_1 &=& \bigoplus_{|\sbmu|=1} \Delta_{\sbmu} ,
\end{eqnarray}
where
\[    |\!\bmu| = \mu_1 + \ldots + \mu_{N-2}\ , \]
gives rise to the circulant structure
\begin{equation}\label{}
    \Sigma_1 = (\oper^{\ot N-1} \ot S) \Sigma_0 \ .
\end{equation}
Another construction goes as follows
\begin{equation}\label{}
    \Xi_{(\alpha|k)} = {\bigoplus_\sbmu}^{(\alpha|k)} \Delta_\sbmu\
    , \ \ \ \alpha=0,1\ , \ \ \ k=1,2,\ldots,N-1\ ,
\end{equation}
where the sum ${\bigoplus}^{(\alpha|k)}$ runs over all $\!\bmu$
with $\mu_k=\alpha$. Note, that $\Xi_{(\alpha|k)}$ displays
circulant structure defined by
\begin{equation}\label{}
    \Xi_{(1|k)} = (\oper^{\ot k} \ot S \ot \oper^{\ot N-k-1} )\,
    \Xi_{(0|k)}\ .
\end{equation}
We shall consider only one scheme with $k=N-1$ and to simplify
notation let us define
\begin{equation}\label{}
    \Xi_\alpha := \Xi_{(\alpha|N-1)}\ , \ \ \ \alpha=0,1\ ,
\end{equation}
which satisfies
\begin{equation}\label{}
    \Xi_1 = (\oper^{\ot N-1} \ot S)\, \Xi_0 \ .
\end{equation}
This very choice has clear interpretation: to define $\Xi_\alpha$
we sum over all $\!\bmu =(\mu_1\ldots\mu_{N-1})$ which represent
binary code for $\alpha$ (mod 2).

Now, let us construct a circulant state
\[  \sigma = \sigma_0 + \sigma_1 \ ,\]
with $\sigma_\alpha$ supported on $\Sigma_\alpha$. It is clear
that
\begin{eqnarray}\label{}
    \sigma_0 &=& \sum_{\sbalpha,\sbbeta} \sum_{i,j=0}^1  a_{\sbalpha i;\sbbeta j} \bigotimes_{k=1}^{N-2}
    e_{\alpha_k\beta_k} \ot  e_{ij} \ot e_{ij} \ , \\
    \sigma_1 &=& \sum_{\sbalpha,\sbbeta} \sum_{i,j=0}^1  b_{\sbalpha i;\sbbeta j} \bigotimes_{k=1}^{N-2}
    e_{\alpha_k\beta_k} \ot  e_{ij} \ot S e_{ij} S^*\ ,
\end{eqnarray}
where $\balpha$ and $\bbeta$ are binary $(N-2)$--vectors with
coordinates $\alpha_k$ and $\beta_k$, respectively, and
$[a_{\sbalpha i;\sbbeta j}]$ and $[b_{\sbalpha i;\sbbeta j}]$ are
$2^{N-1} \times 2^{N-1}$ semi-positive matrices.

Similarly, one constructs a circulant state
\[  \xi = \xi_0 + \xi_1 \ ,\]
with $\xi_\alpha$ supported on $\Xi_\alpha$. It is clear that
\begin{eqnarray}\label{}
    \xi_0 &=& \sum_{\sbalpha,\sbbeta} \sum_{i,j=0}^1
    c_{i\sbalpha;j\sbbeta}\,
    e_{ij} \ot \bigotimes_{k=1}^{N-2} e_{\alpha_k\beta_k} \ot   e_{ij} \ , \\
    \xi_1 &=& \sum_{\sbalpha,\sbbeta} \sum_{i,j=0}^1  d_{i\sbalpha;j\sbbeta}\, e_{ij} \ot \bigotimes_{k=1}^{N-2}
    e_{\alpha_k\beta_k}  \ot S e_{ij} S^*\ ,
\end{eqnarray}
with $2^{N-1} \times 2^{N-1}$ semi-positive  matrices
$[c_{i\sbalpha;j\sbbeta}]$ and $[d_{i\sbalpha i;j\sbbeta }]$.

Now, let us consider  partially transposed $N$-qubit circulant
operators. The corresponding partial transpositions are labeled by
binary $(N-1)$--vectors
\begin{equation}\label{}
    \tau_\sbsigma := \oper \ot \tau^{\sigma_1} \ot \ldots \ot
    \tau^{\sigma_{N-1}}\ .
\end{equation}
Note, that both  $\tau_{\sbsigma} \sigma$ and $\tau_{\sbsigma}\xi$
have exactly the same circulant structure as original $\sigma$ and
$\xi$. One easily finds for the corresponding partial
transpositions:
\[  \tau_{\sbsigma}\xi = \xi_0^{[{\sbsigma}]} + \xi_1^{[{\sbsigma}]} \ ,\]
where $\xi_\alpha^{[{\sbsigma}]}$ are again supported on
$\Xi_\alpha$:
\begin{eqnarray}\label{}
    \xi_0^{[{\sbsigma}]} &=& \sum_{\sbalpha,\sbbeta} \sum_{i,j=0}^1
    c_{i\sbalpha;j\sbbeta}^{[{\sbsigma}]}\,
    e_{ij} \ot \bigotimes_{k=1}^{N-2} e_{\alpha_k\beta_k} \ot   e_{ij} \ , \\
    \xi_1^{[{\sbsigma}]} &=& \sum_{\sbalpha,\sbbeta} \sum_{i,j=0}^1  d_{i\sbalpha;j\sbbeta}^{[{\sbsigma}]}\, e_{ij} \ot \bigotimes_{k=1}^{N-2}
    e_{\alpha_k\beta_k}  \ot S e_{ij} S^*\ ,
\end{eqnarray}
with the new matrices $[c_{i\sbalpha;j\sbbeta}^{[{\sbsigma}]}]$
and $[d_{i\sbalpha i;j\sbbeta }^{[{\sbsigma}]}]$ which are defined
by the following formulae:
\begin{equation}\label{}
    c^{[{\bf 0}]} = c \ , \ \ \ \ d^{[{\bf 0}]} = d\ ,
\end{equation}
and
\begin{equation}\label{}
    c^{[\sbgamma 0]} = \tau_\sbgamma\, c \ , \ \ \ \ d^{[\sbgamma 0]} =
    \tau_\sbgamma\, d \ ,
\end{equation}
where $\!\bgamma$ is binary $(N-2)$--vector and we treat $c$ and
$d$ as a matrices living in the tensor product $ M_2^{\ot N-1}$.
It is therefore clear that $\tau_\sbgamma\, c$ denotes the
corresponding partial transposition of $c$ in the tensor product $
M_2^{\ot N-1}$. Moreover,
\begin{eqnarray}\label{}
 c^{[\sbgamma 1]} &=& c^{[\sbgamma 0]} \circ (\mathbb{I} \ot \widetilde{\mathbb{I}}^{\ot N-2})\,
 +\ d^{[\sbgamma 0]} \circ (S
 \ot \widetilde{\mathbb{I}}^{\ot N-2})\ ,\nonumber \\
d^{[\sbgamma 1]} &=& d^{[\sbgamma 0]} \circ (\mathbb{I} \ot
\widetilde{\mathbb{I}}^{\ot N-2})\, +\ c^{[\sbgamma 0]} \circ (S
 \ot \widetilde{\mathbb{I}}^{\ot N-2})\ .\ \ \ \ \ \ \ \
\end{eqnarray}
\begin{theorem}
A circulant $N$-qubit state $\xi$ is $\bsigma$--{\rm PPT} iff
$c^{[\sbsigma]}$ and $d^{[\sbsigma]}$ are semi-positive matrices.
Similarly, a circulant $N$-qubit state $\sigma$ is $\bsigma$--{\rm
PPT} iff $a^{[\sbsigma]}$ and $b^{[\sbsigma]}$ are semi-positive
matrices.
\end{theorem}

Let us observe that an $N$-qubit circulant state may be easily
reduced to the $(N-L)$-qubit circulant state (with $N-L\geq 2$):
let $l_1,\ldots,l_L$ denote $L$ distinct integers from the set
$\{2,3,\ldots,N-1\}$. Then the partial trace $\mbox{Tr}_{l_1\ldots
l_L}\xi$ defines $(N-L)$-qubit circulant state with new $2^{N-L-1}
\times 2^{N-L-1}$ matrices $c'$ and $d'$ defined by
\begin{equation}\label{}
    c' = \mbox{Tr}_{l_1\ldots l_L}c \ , \ \ \ d' = \mbox{Tr}_{l_1\ldots
    l_L}d\ .
\end{equation}

\section{$N$-qudit state}  \label{N-QUDITS}

Consider now the most general case of $N$ qudits living in
$({\mathbb{C}^d})^{\ot N}$. Again, there are two natural ways to
decompose the corresponding Hilbert space $\mathcal{H}_{d^N}$:
either into $d^{N-1}$ --- $d$-dimensional subspaces, or into $d$
--- $d^{N-1}$--dimensional subspaces.

\subsection{``$\ d^N = d \oplus d \oplus \ldots  \oplus d$''}

Let us introduce a circulant decomposition of $\mathcal{H}_{d^N}$
into $d^{N-1}$ two-dimensional subspaces. Now each integer from
the set $\{0,1,\ldots,d^{N-1}-1\}$ may be represented by a string
of $N-1$ dinaries $(\mu_1\ldots\mu_{N-1})$, i.e. each $\mu_k \in
\{0,1,\ldots,d-1\}$. Let us define $d$-dimensional subspace
\begin{equation}\label{}
 \Delta_{0\ldots 0} =
    \mbox{span}\left\{ e_0 \ot \ldots \ot e_0\,\ldots , e_{d-1} \ot \ldots \ot
    e_{d-1}
    \right\}\ ,
\end{equation}
and for any string of dinaries $(\mu_1\ldots\mu_{N-1})$ define
\begin{equation}\label{}
  \Delta_{\mu_1\ldots\mu_{N-1}} = (\mathbb{I} \ot S^{\mu_1} \ot \ldots  \ot S^{\mu_{N-1}})\Delta_{0\ldots 0}\ .
\end{equation}
Introducing convenient vector notation
\[ \bmu = (\mu_1,\ldots,\mu_{N-1})\ , \]
one has
\begin{equation}\label{}
    \Delta_\sbmu = (\oper \ot S^\sbmu)\Delta_{\bf 0}\ ,
\end{equation}
with
\[ S^\sbmu = S^{\mu_1} \ot \ldots  \ot S^{\mu_{N-1}}\ , \]
and $\Delta_{\bf 0} = \Delta_{0\ldots 0}$. One clearly has
\begin{equation}\label{N-DS-d}
    \mathcal{H}_{d^N} = \bigoplus_\sbmu\, \Delta_\sbmu\ ,
\end{equation}
where the sum runs over all dinary $(N-1)$--vectors $\bmu$.

Now, let us construct a circulant $N$-qubit state $\rho$ based on
(\ref{N-DS-d}):
\begin{equation}\label{}
    \rho = \sum_\sbmu\, \rho_\sbmu\ ,
\end{equation}
where each $\rho_\sbmu$ is supported on $\Delta_\sbmu$. One has
therefore
\begin{equation}\label{}
    \rho_\sbmu = (\oper \ot S^\sbmu) \left[ \sum_{i,j=0}^{d-1}\, x^{(\sbmu)}_{ij} \ e_{ij} \ot
    \ldots \ot e_{ij} \right](\oper \ot S^\sbmu)^* \ ,
\end{equation}
where $[x^{(\sbmu)}]$ are $d\times d$ semi-positive matrices.
Normalization of $\rho$ implies
\begin{equation}\label{}
     \sum_\sbmu\, \mbox{Tr}\,x^{(\sbmu)} = 1\ .
\end{equation}
Now , let us look for the corresponding partial transpositions
$\tau_\sbsigma \rho$ with $\tau_\sbsigma$ introduced in
(\ref{multi-PP}). There is a crucial difference between qubit and
qudit case: for qubits partially transposed state have exactly the
same structure as the original one. It is no longer true for
qudits. It was shown in \cite{CIRCULANT} that partial
transposition gives rise to a new circulant structure governed by
a certain permutation: let $\Pi$ be a $d \times d$ permutation
matrix defined by
\begin{equation}\label{}
    \Pi e_0 = e_0 \ , \ \ \ \ \Pi e_k = e_{d-k} \ ,
\end{equation}
for $k=1,\ldots, d-1$. It turns out \cite{CIRCULANT} that
partially transposed matrix $\tau_\sbsigma \rho$ is related to the
following circulant structure:
\begin{equation}\label{}
    \Delta_\sbmu^{[\sbsigma]} = (\oper \ot S^\sbmu)\Delta_{\bf 0}^{[\sbsigma]}\ ,
\end{equation}
where
\begin{equation}\label{}
    \Delta_{\bf 0}^{[\sbsigma]} = (\oper \ot \Pi^\sbsigma)\Delta_{\bf 0}\ ,
\end{equation}
and
\[ \Pi^\sbsigma = \Pi^{\sigma_1} \ot \ldots  \ot \Pi^{\sigma_{N-1}}\ . \]
One clearly has
\begin{equation}\label{N-DS-d}
    \mathcal{H}_{d^N} = \bigoplus_\sbmu\, \Delta_\sbmu^{[\sbsigma]}\
    ,
\end{equation}
for each binary $(N-1)$--vector $\!\bsigma$.

One finds  therefore the following $\!\bsigma$-circulant structure
for $\tau_\sbsigma \rho$
\begin{equation}\label{}
    \tau_\sbsigma \rho = \sum_\sbmu\, \rho_\sbmu^{[\sbsigma]}\ ,
\end{equation}
with
\begin{widetext}
\begin{equation}\label{}
    \rho_\sbmu^{[\sbsigma]}\, =\, (\oper \ot S^\sbmu)(\oper \ot \Pi^\sbsigma) \left[ \sum_{i,j=0}^{d-1}\, y^{(\sbmu)[\sbsigma]}_{ij} \ e_{ij} \ot
    \ldots \ot e_{ij} \right](\oper \ot \Pi^\sbsigma)^*(\oper \ot S^\sbmu)^* \
    ,
\end{equation}
\end{widetext}
where the new $d \times d$ matrices $y^{(\sbmu)[\sbsigma]}$ are
given by the following formula
\begin{equation}\label{}
    y^{(\sbmu)[\sbsigma]} = \sum_{k=0}^{d-1}\, x^{[\sbmu + k\sbsigma]} \circ ({\Pi \cdot S^k}) \ .
\end{equation}
For $d=2$ one finds $\Pi = \mathbb{I}$ and the above sum reduces
to two terms, only. One therefore recovers (\ref{x-mu-2}).

\begin{theorem}
A circulant state $\rho$ is $\!\bsigma$-{\rm PPT} iff
$y^{(\sbmu)[\sbsigma]}$ are semi-positive for all dinary $\bmu$.
\end{theorem}

{\it Examples.} 1. Generalized GHZ state

\begin{equation}\label{}
    |{{\rm GHZ}}\> = \frac{1}{\sqrt{d}} \, \sum_{k=0}^{d-1}\, e_k \ot \ldots \ot e_k\ ,
\end{equation}
does belong to circulant class which is easily seen from the
corresponding density matrix with
\begin{equation}\label{}
    x^{(\sbmu)} =\, \delta_{\sbmu,{\bf 0}}\,
    \widetilde{\mathbb{I}} \ ,
\end{equation}
where
\begin{equation}\label{}
 \widetilde{\mathbb{I}} = \sum_{\alpha=0}^{d-1} \, S^\alpha\ ,
\end{equation}
generalizes (\ref{tilde-I-2}), that is,
$\widetilde{\mathbb{I}}_{ij}=1$ for all $i,j=0,1,\ldots,d-1$.

2. Generalized Bell states: the following $d^N$ vectors
\begin{equation}\label{}
    \psi_{\alpha\sbnu} = ( \Omega^\alpha \ot S^\sbnu) |\rm{GHZ}\> \ ,
\end{equation}
where the phase operator $\Omega$ is defined via
\begin{equation}\label{}
    \Omega e_k = \omega^k e_k \ , \ \ \ \ k=0,1,\ldots,d-1\
    ,
\end{equation}
with $\omega = e^{2\pi i /d}$, define a circulant state for any
 $\alpha=0,1,\ldots,d-1$ and  arbitrary dinary $(N-1)$--vector $\!\bnu$.
 These are $N$-qudit generalization of $N$-qubit
Bell states (\ref{2N-Bell}).  Note that
\begin{equation}\label{}
 \psi_{\alpha\sbnu} \in \Delta_{\sbnu}\ ,
\end{equation}
and the corresponding matrices $x^{(\sbmu)}$ read as follows
\begin{equation}\label{}
    x^{(\sbmu)} =\, \delta_{\sbmu,\sbnu}\,
    \widetilde{\Omega} \ ,
\end{equation}
where the $d\times d$ matrix $\widetilde{\Omega}$ is defined by
\begin{equation}\label{}
 \widetilde{\Omega}_{ij} = \omega^{j-i}\ ,
\end{equation}
and generalizes a $2 \times 2$ matrix from (\ref{2x2}).

 3. Generalized $N$-qudit isotropic state

\begin{equation}\label{}
    \rho = \frac{1-s}{d^N}\,\mathbb{I}^{\ot N} + s \,|\rm{GHZ}\>\<\rm{GHZ}|
\end{equation}
with $s\in [-1/(d^N-1),1]$. One finds for $x^{(\sbmu)}$ matrices
\begin{equation}\label{}
    x^{({\bf 0})} = \frac{1}{d^N} \left( \begin{array}{cc} 1+(d^N-1)s & d^{N-1}s \\ d^{N-1}s &
    1+(d^N-1)s \end{array} \right)\ ,
\end{equation}
and
\begin{equation}\label{}
    x^{(\sbmu)} =  \frac{1-s}{d^N}\,\mathbb{I}\
    ,
\end{equation}
for $\bmu \neq {\bf 0}$. The only nontrivial PPT condition comes
from the positivity of
\begin{equation}\label{}
 \left( \begin{array}{cc} 1-s & d^{N-1}s \\  d^{N-1}s & 1-s  \end{array} \right)\ ,
\end{equation}
which implies
\[ s \leq \frac{1}{d^{N-1} +1}\ .\]
The above condition guaranties full $N$-separability of $\rho$
\cite{Art}.

\subsection{``$\ d^N = d^{N-1} \oplus d^{N-1}$''}

There are several ways to construct $d^{N-1}$--dimensional
circulant decompositions of $N$-qudit Hilbert space out of
$d$-dimensional mutually orthogonal spaces $\Delta_{\sbmu}$. The
following choice
\begin{eqnarray}\label{}
    \Sigma_\alpha = \bigoplus_{|\sbmu|=\alpha} \Delta_{\sbmu} \ , \ \ \ \ \
    \alpha=0,1,\ldots,d-1
\end{eqnarray}
gives rise to the circulant structure
\begin{equation}\label{}
    \Sigma_\alpha = (\oper^{\ot N-1} \ot S^\alpha) \Sigma_0 \ .
\end{equation}
Another construction goes as follows
\begin{equation}\label{}
    \Xi_{(\alpha|k)} = {\bigoplus_\sbmu}^{(\alpha|k)} \Delta_\sbmu\
    ,
\end{equation}
for $\alpha = 0,1,\ldots,d-1$ and $k=1,2,\ldots,N-1$. In the above
formula  the sum ${\bigoplus}^{(\alpha|k)}$ runs over all $\!\bmu$
with $\mu_k=\alpha$. Note, that $\Xi_{(\alpha|k)}$ displays
circulant structure defined by
\begin{equation}\label{}
    \Xi_{(\alpha|k)} = (\oper^{\ot k} \ot S^\alpha \ot \oper^{\ot N-k-1} )\,
    \Xi_{(0|k)}\ .
\end{equation}
We shall consider only one scheme with $k=N-1$ and to simplify
notation let us define
\begin{equation}\label{}
    \Xi_\alpha := \Xi_{(\alpha|N-1)}\ , \ \ \ \alpha=0,1,\ldots,d-1\ ,
\end{equation}
which satisfies
\begin{equation}\label{}
    \Xi_\alpha = (\oper^{\ot N-1} \ot S^\alpha)\, \Xi_0 \ .
\end{equation}
This very choice has clear interpretation: to define $\Xi_\alpha$
we sum over all $\!\bmu =(\mu_1\ldots\mu_{N-1})$ which represent
dinary code for $\alpha$ (mod $d$).

Now, let us construct a circulant state
\[  \sigma = \sum_{\alpha=0}^{d-1}\, \sigma_\alpha \ ,\]
with $\sigma_\alpha$ supported on $\Sigma_\alpha$. It is clear
that
\begin{eqnarray}\label{}
    \sigma_\alpha =  \sum_{\sbalpha,\sbbeta} \sum_{i,j=0}^{d-1}  a^{(\alpha)}_{\sbalpha i;\sbbeta j} \bigotimes_{k=1}^{N-2}
    e_{\alpha_k\beta_k} \ot  e_{ij} \ot S ^\alpha e_{ij} S^{\alpha *} \ ,
\end{eqnarray}
where $\balpha$ and $\bbeta$ are dinary $(N-2)$--vectors with
coordinates $\alpha_k$ and $\beta_k$, respectively, and
$[a^{(\alpha)}_{\sbalpha i;\sbbeta j}]$ is a set of $d^{N-1}
\times d^{N-1}$ semi-positive matrices. This set generalizes two
matrices $a=a^{(0)}$ and $b=a^{(1)}$ in the qubit case, i.e.
$d=2$.

Similarly, one constructs a circulant state
\[  \xi = \sum_{\alpha=0}^{d-1} \, \xi_\alpha \ ,\]
with $\xi_\alpha$ supported on $\Xi_\alpha$. It is clear that
\begin{eqnarray}\label{}
    \xi_\alpha = \sum_{\sbalpha,\sbbeta} \sum_{i,j=0}^{d-1}
    c^{(\alpha)}_{i\sbalpha;j\sbbeta}\,
    e_{ij} \ot \bigotimes_{k=1}^{N-2} e_{\alpha_k\beta_k} \ot  S^\alpha e_{ij} S^{\alpha *} \ ,
\end{eqnarray}
with $d^{N-1} \times d^{N-1}$ semi-positive  matrices
$[c^{(\alpha)}_{i\sbalpha;j\sbbeta}]$.

Now, each partial transposition $\tau_\sbsigma$ gives rise to the
new circulant structure: either
\begin{eqnarray}\label{}
    \Sigma_\alpha^{[\sbsigma]} = \bigoplus_{|\sbmu|=\alpha} \Delta_{\sbmu}^{[\sbsigma]} \ , \ \ \ \ \
    \alpha=0,1,\ldots,d-1
\end{eqnarray}
with the cyclic property
\begin{equation}\label{}
    \Sigma_\alpha^{[\sbsigma]} = (\oper^{\ot N-1} \ot S^\alpha) \Sigma_0^{[\sbsigma]} \
    ,
\end{equation}
or
\begin{equation}\label{}
    \Xi_\alpha^{[\sbsigma]} = {\bigoplus_\sbmu}^{(\alpha|N-1)} \Delta_\sbmu^{[\sbsigma]}\
    ,
\end{equation}
with the same property, that is,
\begin{equation}\label{}
    \Xi_\alpha^{[\sbsigma]} = (\oper^{\ot N-1} \ot S^\alpha) \Xi_0^{[\sbsigma]} \
    .
\end{equation}
One easily finds for the corresponding partial transpositions:
\[  \tau_{\sbsigma}\xi = \sum_{\alpha=0}^{d-1}\, \xi_0^{[{\sbsigma}]}  \ ,\]
where $\xi_\alpha^{[{\sbsigma}]}$ are  supported on
$\Xi_\alpha^{[{\sbsigma}]}$:
\begin{eqnarray}\label{}
    \xi_\alpha^{[{\sbsigma}]} & =& \sum_{\sbalpha,\sbbeta}
    \sum_{i,j=0}^{d-1}
    a^{(\alpha)[\sbsigma]}_{i\sbalpha;j\sbbeta}\\ \nonumber &&
    e_{ij} \ot \bigotimes_{k=1}^{N-2} e_{\alpha_k\beta_k} \ot  S^\alpha \Pi e_{ij} \Pi^* S^{\alpha *} \
    ,
\end{eqnarray}
with new set of matrices
$[c_{i\sbalpha;j\sbbeta}^{(\alpha)[{\sbsigma}]}]$  which are
defined by the following formulae:
\begin{equation}\label{}
    c^{(\alpha)[{\bf 0}]} = c^{(\alpha)} \ ,
\end{equation}
and
\begin{equation}\label{}
    c^{(\alpha)[\sbgamma 0]} = \tau_\sbgamma\, c^{(\alpha)} \ ,
\end{equation}
where $\!\bgamma$ is binary $(N-2)$--vector and we treat
$c^{(\alpha)}$  matrices living in the tensor product $ M_d^{\ot
N-1}$. It is therefore clear that $\tau_\sbgamma\, c^{(\alpha)}$
denotes the corresponding partial transposition of $c^{(\alpha)}$
in the tensor product $ M_d^{\ot N-1}$. Moreover,
\begin{eqnarray}\label{}
 c^{(\alpha)[\sbgamma 1]} &=& \sum_{\beta=0}^{d-1} c^{(\alpha+\beta)[\sbgamma 0]} \circ (\Pi S^\beta \ot \widetilde{\mathbb{I}}^{\ot N-2})
  \ .
\end{eqnarray}
\begin{theorem}
A circulant $N$-qudit state $\xi$ is $\bsigma$--{\rm PPT} iff
$c^{(\alpha)[\sbsigma]}$  are semi-positive matrices for
$\alpha=0,1,\ldots,d-1$.
\end{theorem}

\section{Conclusions}

We have constructed a large class of PPT states  which
 correspond to circular decompositions of $\mathbb{C}^d \ot \ldots
 \ot \mathbb{C}^d$ into direct sums of $d$-- and $d^{N-1}$--dimensional subspaces. This
class generalizes bipartite circulant states introduced in
\cite{CIRCULANT}. It contains several known examples from the
literature and produces a highly nontrivial family of new states.

There are several open problems related to this new class: the
basic question is how to detect entanglement within this class of
multipartite states. One may expect that there is special class of
entanglement witnesses which are sensitive to entanglement encoded
into circular decompositions, that is, circulant Hermitian
operators $W \in M_d^{\ot N}$ such that
\begin{equation}\label{}
    \mbox{Tr}(W \rho_1 \ot \ldots \ot \rho_N) \geq 0 \ ,
\end{equation}
for all product states $\rho_1 \ot \ldots \ot \rho_N$, and
\begin{equation}\label{}
    \mbox{Tr}(W \xi) < 0 \ ,
\end{equation}
for some circulant state $\xi$. It is interesting to explore the
possibility of other decompositions leading to new classes of
multipartite states. Let us note, that so called W state of 3
qubits (\ref{W-state}) does not belong to our class. Another
important family of states which does not fit circulant class was
introduced in \cite{Werner3}: these are $N$-qudit states
satisfying
\begin{equation}\label{UUU}
    U \ot \ldots \ot U \, \rho =  \rho\, U \ot \ldots \ot U\ ,
\end{equation}
for all unitaries $U \in U(d)$. For $N=2$ it reduces to the Werner
state \cite{Werner} which belongs to bipartite circulant class.
However, it is easy to check that for $N\geq 3$ states satisfying
(\ref{UUU}) are not circulant. One may expect the existence of
other characteristic decompositions which are responsible for the
structure of symmetric states governed by (\ref{UUU}). Anyway,
multipartite circulant states introduced in this paper may shed
new light on the more general investigation of multipartite
entanglement.

\acknowledgments  This work was partially supported by the Polish
State Committee for Scientific Research.


\begin{thebibliography}{1} \bibliographystyle{plain}


\bibitem{QIT} M. A. Nielsen and I. L. Chuang, {\it Quantum computation
and quantum information}, Cambridge University Press, Cambridge,
2000.

\bibitem{Horodecki-review} R. Horodecki, P. Horodecki, M. Horodecki and K.
Horodecki, {\it Quantum entanglement}, arXiv: quant-ph/0702225.


\bibitem{Peres} A. Peres, Phys. Rev. Lett. {\bf 77}, 1413 (1996).

\bibitem{PPT} P. Horodecki, Phys. Lett. A {\bf 232}, 333 (1997).

\bibitem{CIRCULANT} D. Chru\'sci\'nski and A. Kossakowski, Phys.
Rev. A. {\bf 76}, 032308   (2007).

\bibitem{MULTI1} A. Miyake and H-J. Briegel, Phys. Rev. Lett. {\bf 95}, 220501
(2005).

\bibitem{MULTI2}  A.C. Doherty, P.A. Parrilo and  F.M. Spedalieri,
 Phys. Rev. A, Vol. {\bf 71}, 032333 (2005).

\bibitem{MULTI3}  G. Toth and O. Guehne, Phys. Rev. Lett. {\bf 94}, 060501
(2005).

\bibitem{MULTI4} M. Bourennane, M. Eibl, Ch. Kurtsiefer, S.
Gaertner, H. Weinfurter, O. Guehne, P. Hyllus, D. Bruss, M.
Lewenstein and A. Sanpera,  Phys. Rev. Lett. {\bf 92}, 087902
(2004).

\bibitem{MULTI5} A. Acin,  Phys. Rev.
Lett. {\bf 88}, 027901 (2002)

\bibitem{MULTI6} W. D\"ur, J. I. Cirac and R. Tarrach, Phys. Rev.
Lett. {\bf 83}, 3562 (1999); W. D¨ur and J.I. Cirac, Phys. Rev. A
{\bf 61}, 042314 (2000).

\bibitem{Werner3} T. Eggeling  and R.F. Werner, Werner, Phys. Rev. A {\bf 63}, 042111 (2001).

\bibitem{I&II} D. Chru\'sci\'nski and A. Kossakowski,  Phys.
Rev. A. {\bf 73}, 062313  (2006); Phys. Rev. A. {\bf 73}, 062314
(2006).

\bibitem{Werner} R.F. Werner, Phys. Rev. A {\bf 40}, 4277 (1989).

\bibitem{ISO} M. Horodecki and P. Horodecki, Phys. Rev. {\bf A} 59, 4206
(1999).

\bibitem{Werner2} K.G.H. Vollbrecht and R.F. Werner, Phys. Rev. A {\bf 64}, 062307 (2001).

\bibitem{GHZ} D.M. Greenberger, M. Horne and A. Zelinger, in
{\it Bell's theorem , Quantum Theory and Conceptions of the
Universe}, edited by M. Kafatos (Kluwer Academic, Dordrecht, The
Netherlands, 1989), pp. 69.

\bibitem{Art} A. Pittenger and M. Rubin, Optics Communications,
{\bf 179}, 447 (2000).

\bibitem{Art2} A. Pittenger and M. Rubin, Phys. Rev. A {\bf 67}, 012327 (2003).

\bibitem{Art3} A. Pittenger and M. Rubin, Phys. Rev. A {\bf 62}, 042306 (2000).

\end{thebibliography}
\end{document}